\documentclass[aps,prd,floatfix,nofootinbib,amssymb,preprintnumbers,preprint]{revtex4}
\usepackage{epsf}
\usepackage{graphicx}

\def\lQ{\Lambda_{\rm QCD}}
\newcommand{\nn}{\nonumber}
\newcommand{\be}{\begin{equation}}
\newcommand{\ee}{\end{equation}}
\newcommand{\bea}{\begin{eqnarray}}
\newcommand{\eea}{\end{eqnarray}}

\def\als{\alpha_{\rm s}}
\def\siml{{\ \lower-1.2pt\vbox{\hbox{\rlap{$<$}\lower6pt\vbox{\hbox{$\sim$}}}}\ }} 

\newcommand{\MS}{\overline{\rm MS}}
\newcommand{\RS}{\rm RS}

\newcommand{\OS}{\rm OS}

\newcommand{\PV}{\rm PV}

\def\simg{{\ \lower-1.2pt\vbox{\hbox{\rlap{$>$}\lower6pt\vbox{\hbox{$\sim$}}}}\ }}
\def\siml{{\ \lower-1.2pt\vbox{\hbox{\rlap{$<$}\lower6pt\vbox{\hbox{$\sim$}}}}\ }} 

\begin{document}
\vskip 1truecm
\title{Comment on ``The MSR Mass and the ${\cal O}(\Lambda_{\rm QCD})$ Renormalon Sum Rule''}
\author {Antonio Pineda}
\address{Grup de F\'\i sica Te\`orica, Dept. F\'\i sica and IFAE-BIST, Universitat Aut\`onoma de Barcelona,\\ 
E-08193 Bellaterra (Barcelona), Spain}

\begin{abstract}
In this note we confront the research conducted in \cite{Hoang:2017suc} with results previously obtained by the author, and critically examine some of their claimed findings. 
\end{abstract}

\maketitle      

\pagenumbering{arabic}

\section{Introduction}

Recently the paper entitled ``The MSR Mass and the ${\cal O}(\Lambda_{\rm QCD})$ Renormalon Sum Rule'' \cite{Hoang:2017suc} appeared in the arxives. There are aspects of this paper, such as the existence of renormalon sum rules for the pole mass, or explicit resummations of renormalon-associated logarithms for renormalon-free masses, that existed prior to the work by the authors. It is the aim of this note to clarify some aspects of this.

In order to proceed with the discussion we need to introduce some preliminary material first, which we mainly copy from Ref.~\cite{Pineda:2001zq}. The pole mass and the 
$\MS$ renormalized mass are related by the following perturbative series
\be
\label{series}
m_{\OS} = m_{\MS} + \sum_{n=0}^\infty r_n(\nu) \als^{n+1}(\nu)\,. 
\ee
We then define the Borel transform 
\be\label{borel}
m_{\OS} = m_{\MS} + \int\limits_0^\infty\mbox{d} t \,e^{-t/\als}
\,B[m_{\OS}](t)
\,,
\qquad B[m_{\OS}](t)\equiv \sum_{n=0}^\infty 
r_n \frac{t^n}{n!} . 
\ee
We also define 
$$
\nu {d \als \over d
\nu}=-2\als\left\{\beta_0{\als \over 4 \pi}+\beta_1\left({\als \over 4
\pi}\right)^2 + \cdots\right\}
.$$  
The behavior of the perturbative expansion of
Eq. (\ref{series}) at large
orders is dictated by the closest singularity to the origin of its
Borel transform, which happens to be located at
$t=2\pi/\beta_0$ \cite{Bigi:1994em,Beneke:1994sw}.
Being more precise, the behavior of the Borel transform near the
closest singularity at the origin reads (we define $u={\beta_0 t \over 4 \pi}$)
\be
B[m_{\OS}](t(u))=N_m\nu {1 \over
(1-2u)^{1+b}}\left(1+c_1(1-2u)+c_2(1-2u)^2+\cdots \right)+({\rm
analytic\; term}),
\ee
where by {\it analytic term}, we mean a function expected to be
analytic up to the next renormalon. This dictates the behavior of the perturbative expansion at large orders to be 
\be\label{generalm}
r_n \stackrel{n\rightarrow\infty}{=} 
r_n^{as}=N_m\,\nu\,\left({\beta_0 \over 2\pi}\right)^n
\,{\Gamma(n+1+b) \over
\Gamma(1+b)}
\left(
1+\frac{b}{(n+b)}c_1+\frac{b(b-1)}{(n+b)(n+b-1)}c_2+ \cdots
\right).
\ee
The coefficients $b$ and $c_1$ were computed in \cite{Beneke:1994rs}, and $c_2$ in \cite{Beneke:1998ui}, \cite{Pineda:2001zq} . The latter reference corrects some missprints for the analogous expression $s_2$.\footnote{Incidentally, this has gone unnoticed by the authors of Ref.~ \cite{Hoang:2017suc}, who claim agreement for the $s_2$ expression in \cite{Beneke:1998ui}.} They read 
\be
b={\beta_1 \over 2\beta_0^2}\,,
\qquad
c_1={1 \over 4\,b\beta_0^3}\left({\beta_1^2 \over \beta_0}-\beta_2\right)
\ee
and 
\be
c_2={1 \over b(b-1)}
{\beta_1^4 + 4 \beta_0^3 \beta_1 \beta_2 - 2 \beta_0 \beta_1^2 \beta_2 + 
   \beta_0^2 (-2 \beta_1^3 + \beta_2^2) - 2 \beta_0^4 \beta_3 
\over 32 \beta_0^8}
\,.
\ee

Obviously the very same existence of the pole mass renormalon relies on a nonzero value of $N_m$.

\section{Sum rules for $N_m$}

The authors of \cite{Hoang:2017suc} make an strong case about the derivation of what they name ``the'' sum rule for the determination of  $N_m$. Well, we want to emphasize that sum rules for $N_m$ existed before. In Ref. \cite{Pineda:2001zq} the following ``sum rule'' was used for the determination of $N_m$\footnote{Adapting the discussion of Ref.~\cite{Lee:1996yk} on the gluon condensate to the pole mass where the convergence is much better.}:
\bea
\label{Nmsumrule}
N_m&=&\frac{1}{\nu}(1-2u)^{1+b}B[m_{\OS}](t(u))\Bigg|_{u=1/2}
\\
&=&
\frac{1}{\nu}\sum_{m,n=0}^{\infty}\frac{\Gamma(2+b)(-1)^m r_n(\nu)}{\Gamma(m+1)\Gamma(n+1)\Gamma(2+b-m)}\left(\frac{2\pi}{\beta_0}\right)^n
.
\nn
\eea
This gave the first numerical determination of $N_m$ beyond the large $\beta_0$ limit \cite{Beneke:1994sw}.

\section{Numerical determination of $N_m$}

\begin{figure}
\hspace{-3cm}
\centerline{\includegraphics[width=0.5\columnwidth]{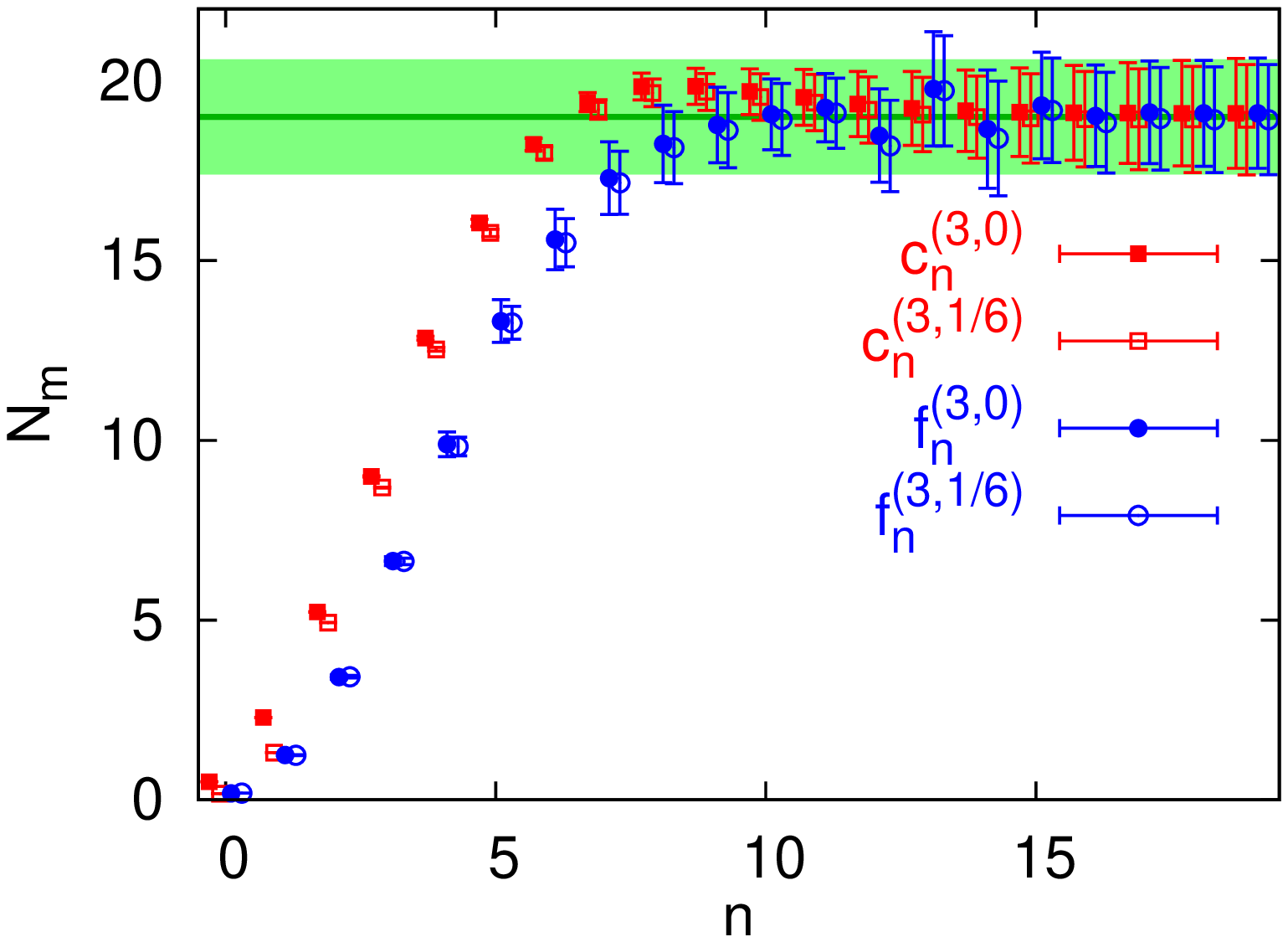}
\includegraphics[width=0.5\textwidth]{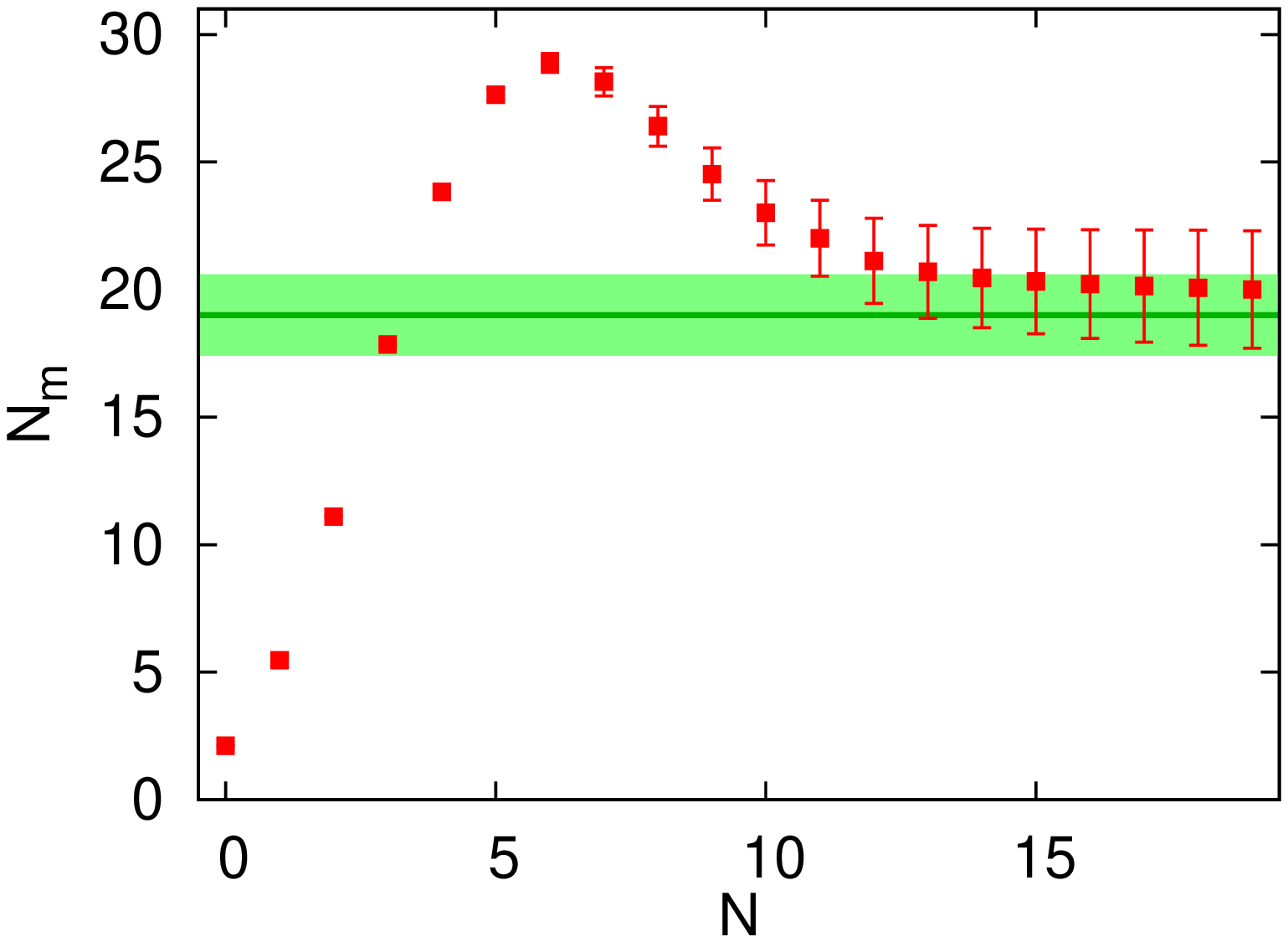}}
\vspace{-0.5cm}
\caption{{\it From Ref.~\cite{Bali:2013pla}. Determination of $N_m$ in the lattice scheme. The left figure is the determination of $N_m$ via Eq.~(\ref{Nmratio}), from the
coefficients $c_n^{(3,0)},c_n^{(3,1/6)},f_n^{(3,0)}$ and $f_n^{(3,1/6)}$. The right figure is the determination of 
$N_m$ via Eq.~(\ref{Nmsumrule}) from the coefficients $c_n^{(3,0)}$.
The horizontal band is the error estimate. For extra details see \cite{Bali:2013pla}.
}
\label{fig:Nm(n)}}
\end{figure}

The authors of \cite{Hoang:2017suc} estimate the error for $N_m$ based on the scale variation. Yet, the typical size of the difference between consecutive orders in their Fig. 3 (not clear if it refers to $n_l=4$ or 5), may indicate larger errors. Theoretically, there is not a clear reason, a priori, to prefer one sum rule versus another, nor to know what the optimal method to determine $N_m$ is. This has to be studied by carefully comparing different methods (and also different observables) under similar conditions\footnote{Some may even have doubts about the existence of the pole mass renormalon, as only few terms of the perturbative expansion are known. This motivated dedicated lattice simulations to prove the existence of the pole mass renormalon beyond any reasonable doubt by computing the static energy of a heavy quark to order 20 \cite{Bauer:2011ws,Bali:2013pla}.}. So, obviously here there is room for dedicated studies. 
 For instance, in Ref.~\cite{Bali:2013pla} the sum rule method of Eq.~(\ref{Nmsumrule}) was confronted with an alternative method:
\be
\label{Nmratio}
N_m=\lim_{n\rightarrow\infty}\frac{r_n}{(r_n^{as}/N_m)}
\,.
\ee
 It was shown (using the static energy of a heavy quark, replacing $r_n \rightarrow c_n$, $f_n$) that while both Eqs.~(\ref{Nmratio}) and (\ref{Nmsumrule}) converge to the expected number they are not equally efficient in doing so, see Fig.~\ref{fig:Nm(n)}. A similar conclusion but in weaker terms was reached in \cite{Ayala:2014yxa}.

\section{Renormalization group improvement of renormalon associated effects}

Finally, the second main point of Ref.~\cite{Hoang:2017suc} is the resummation of logarithms associated to the renormalon. For instance, 
the authors state

``the conceptual implications of R-evolution and its connection to the $O(\Lambda_{QCD})$ renormalon
problem in the perturbative relations between short-distance masses and the pole mass were
first studied systematically in Ref. [44].'' \\
and alike. Well, the resummation of logarithms associated with the pole mass renormalon was already computed in Ref. \cite{Bali:2003jq} 
(see also \cite{Campanario:2005np}) in terms of incomplete Gamma functions and directly related to $N_m$. In those references the renormalon subtracted mass (RS) was used. It is defined by
\be
\label{mrsvsmpole}
m_{\RS}(\nu_f)=m_{\OS}-\delta m_{\RS}(\nu_f)=m_{\MS} + \sum_{n=0}^\infty r^{\RS}_n\als^{n+1}(\nu)
\,,
\ee
where $r^{\RS}_n=r^{\RS}_n(m_{\MS},\nu,\nu_f)$ and  
\be
\delta m_{\RS}(\nu_f)=\sum_{n=n_{min}}^\infty  N_m\,\nu_f\,\left({\beta_0 \over
2\pi}\right
)^n \als^{n+1}(\nu_f)\,\sum_{k=0}^\infty c_k{\Gamma(n+1+b-k) \over
\Gamma(1+b-k)}
\,.
\ee
In what follows we take $n_{min}=1$ so that we can follow almost verbatim the discussion in Ref. \cite{Bali:2003jq}, 
but other options are possible. 

The running of the RS mass with $\nu_f$ is renormalon-free.
Therefore, it
can be described by a convergent expansion in
perturbation theory. Nevertheless, in order to achieve the renormalon
cancellation, the same scale $\nu$ has to be used in the perturbative
expansion. This produces large logs if the scales $\nu_f$ and
$\nu_f'$ are widely separated and, eventually, some errors, if one
works to finite order in perturbation theory. In the RS scheme, there
exists a solution to this problem. Even though $\delta m_{\RS}(\nu_f)$
suffers from the renormalon ambiguity, the difference 
\be
m_{\RS}(\nu_f)- m_{\RS}(\nu_f')= \delta m_{\RS}(\nu_f)-\delta m_{\RS}(\nu_f')
\ee 
is renormalon-free.
We can perform a resummation of $\delta m_{\RS}(\nu_f)$
with any prescription to avoid the singularity in the Borel
plane since it will cancel in the difference. The
Principal Value (PV) prescription yields
\be
\label{PV}
\delta m_{\RS}^{\rm PV}(\nu_f)= N_m\nu_f\als(\nu_f)\sum_{s=0}^\infty
c_s 
\left[
D_{b-s}\left( -{2\pi \over \beta_0\als(\nu_f)}\right) 
-1
\right]
\,,  
\ee 
where (a typo in \cite{Bali:2003jq} was corrected in \cite{Campanario:2005np})
\be
\label{DPV}
D_{b}(-x)=-xe^{-x}x^b\cos(\pi b)\Gamma(-b)+x
(-x)^{b}e^{-x}\left[\Gamma(-b)-\Gamma(-b,-x)\right]
\,,
\ee
and,
\be
\Gamma(b,x)=\int_x^\infty\! dt\, t^{b - 1} e^{-t}\,,
\ee
denotes the incomplete $\Gamma$ function.

\begin{figure}
\includegraphics[width=0.8\columnwidth]{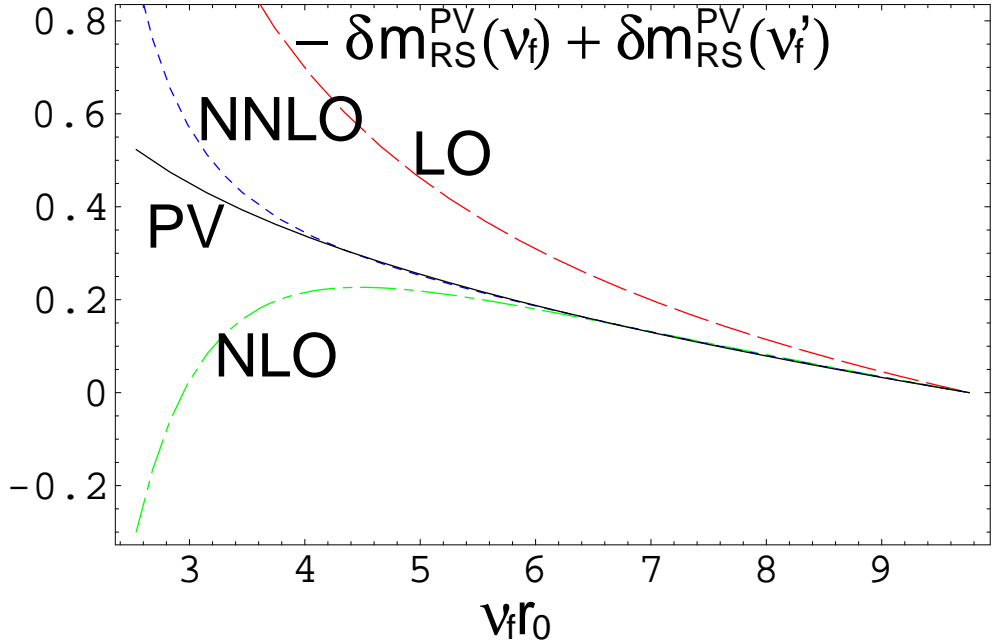}
\caption{\label{PVvsFO} \it From Ref.~\cite{Bali:2003jq}. 
$-\delta m_{\RS}(\nu_f)+\delta m_{\RS}(\nu_f')$
at LO (dashed line), NLO (dashed-dotted line) and NNLO (dotted line)
in perturbation
theory versus the Principal Value result (solid line). We take
$\nu_f'=9.76\, r_0^{-1}$. }
\end{figure}

The first term in Eq. (\ref{DPV}) corresponds to $\Lambda_{\MS}$,
once introduced in the sum of Eq. (\ref{PV}). It cancels from
the combination,
$\delta m_{\RS}^{\rm PV}(\nu_f)-\delta m_{\RS}^{\PV}(\nu_f')$. The sum of Eq.\ (\ref{PV})
represents softer and softer singularities in the Borel
plane. Therefore, we expect the sum to converge for the difference $\delta
m_{\RS}^{\PV}(\nu_f)-\delta m_{\RS}^{\PV}(\nu_f')$. Since the first three terms are
known one can check that this actually happens.
We can also compare $-\delta m_{\RS}^{\rm PV}(\nu_f)+\delta
m_{\RS}^{\rm PV}(\nu_f')$ with the corresponding difference, calculated at
finite order in perturbation theory. 
We depict this comparison in Fig.\ \ref{PVvsFO} (for $n_f=0$ and $\lQ=0.602\;r_0^{-1}$), where we take
$\nu=\nu_f$ to minimise one of the logs. We see how the finite order
results approach the PV curve. 

We let the reader to compare the formulas above to those of Ref.~\cite{Hoang:2017suc}. In this context it is quite revealing that Ref. \cite{Bali:2003jq} is not in the list of 114 references of v1 of Ref.~\cite{Hoang:2017suc}.

\section{Comments to v2 of Ref.~\cite{Hoang:2017suc}}

A new version of \cite{Hoang:2017suc} has recently appeared in the arxives. We would like to state that the modifications made in that reference do not change the main two points of this note:
\begin{enumerate}
\item
Sum rules for $N_m$ existed before.
\item
Pole mass renormalon-associated logarithms have been resummed before.
\end{enumerate}
Looking to v2 of \cite{Hoang:2017suc} the impression the reader will get is otherwise. 

\medskip

We profit to make some extra comments to the modifications made in  v2 of \cite{Hoang:2017suc}, which are potentially missleading. First we stress that obtaining the sum rule is kind of trivial. Not a big deal. One only Taylor expands $D(u)\equiv\frac{1}{\nu}(1-2u)^{1+b}B[m_{\OS}](t(u))$ around $u=1/2$. Eq.  (\ref{Nmsumrule}) then corresponds to (see Refs.~\cite{Pineda:2001zq,Bali:2013pla}):  
\be
N_m=\lim_{N\rightarrow \infty}\sum_{n=0}^{N}\frac{1}{n!} \frac{ d^n D_m}{d u^n}\Bigg|_{u=0} \left(\frac{1}{2}\right)^n
\,.
\ee

A more serious misstatement in  v2 of \cite{Hoang:2017suc} is their push for a unique sum rule different from the one originally proposed in Ref. \cite{Pineda:2001zq}. 
This requires a more detailed discussion. Although it is not clear in their presentation, the physical input introduced in \cite{Hoang:2017suc} is equivalent to applying the operator product expansion (OPE), otherwise there is no cancellation of the renormalon. Stated in a different way: if one changes the power of R in Eq. (2.3) in v2 of 
\cite{Hoang:2017suc}, and/or multiply it by a function of $\ln R$, one can fictitiously generate renormalons with arbitrarily different dimensionality and with different branch cut structure. Therefore, hidden in Eq. (2.3) of v2 of \cite{Hoang:2017suc} is the knowledge of the OPE structure of the pole mass. This is exactly the input used in \cite{Pineda:2001zq} (which comes back to simply the fact that the $B$ meson mass, $M_B=m_{OS}+\bar \Lambda$, is renormalon free). Indeed this was originally quantified in Ref.~\cite{Beneke:1994rs} using renormalization group techniques, which is equivalent to say that the dimension of the higher twist is 1 and its Wilson coefficient is 1. This makes evident that the physical input is equal. And actually the analytic structure of the Borel transform of the pole mass used by the authors is the same. Still, the authors claim for a different sum rule, and that the difference do not vanish. Whereas it is not completely clear to the author how they generate the difference, it seems to be related with analytic terms of $B[m_{\OS}]$ with a radius of convergence bigger than $u=1/2$.\footnote{If it were not so, it would imply that the analytic structure of $B[m_{\OS}]$ used in \cite{Hoang:2017suc} would be different from the one predicted by the OPE.} We note that such functions, when multiplied by $(1-2u)^{1+b}$ converge to zero when considering their Taylor expansion around $u=1/2$. Still the authors of \cite{Hoang:2017suc} argue for a nonzero difference. They quantify this difference stating that higher order  beta coefficients are compulsory to get $N_m$. 
In such a claim one has a clear smoking gun to falsify their conclusion. The right figure in Fig.~\ref{fig:Nm(n)} is a clear counterexample of that statement. It converges to $N_m$ and does not use other beta coefficients than $\beta_0$ and $\beta_1$.

\medskip

Though not the purpose of this note, let us make some final considerations. 

1) In \cite{Bali:2013pla,Ayala:2014yxa} we have observed (numerically) that the determination of $N_m$ through the ratio Eq.~(\ref{Nmratio}) gives better numerical results than the sum rule.
In this respect we would also like to mention that this method allows for a parametric estimate of the error. The error associated to the log-independent terms is of the order $1/n^{k+1}$ where $k$ is the power of the last known term of the $1/n$ expansion of asymptotic behavior of $r_n$ (constrained by the knowledge of the coefficients of the beta-function for the case of the pole mass, and by the knowledge of the Wilson coefficients in more general cases). Log-related terms 
($\sim \ln \nu/m$) are exponentially supressed $\sim e^{-\# n}$ as they are related, at most, with subleading renormalons. The importance of these errors numerics will tell (provided one has enough terms to check). 

2) Thinking in top physics, in Ref. \cite{Hoang:2017suc} a lot emphasis has been put in determinations of $N_m$ for $n_f=5$ . We would like to stress that in the long term, even in top physics, the most relevant determination will be with  $n_f=3$ (and $n_f=0$ if one is interested in gluodynamics or direct comparison with quenched lattice simulations). The reason is that if one goes to high enough orders in perturbation theory the bottom and charm decouple. And we want to go to high enough orders in perturbation theory if we want to see renormalon associated effects, and achieve precision of the order or smaller than $\lQ$. 
At those high orders the renormalon will see three active  light flavours and not four or five.

3) In v2 of Ref. \cite{Hoang:2017suc} the authors compare with the $n_f=0$ determination of $N_m$ from lattice simulations of Ref.~\cite{Bauer:2011ws}. The most up-to-date determination was obtained in Ref.~\cite{Bali:2013qla} and reads $N_m=0.620(35)$.

\begin{acknowledgments}
This work was supported in part by the Spanish grants FPA2014-55613-P, FPA2013-
43425-P, SEV-2012-0234, and the Catalan grant SGR2014-1450.
\end{acknowledgments}



\begin{thebibliography}{99}

\bibitem{Hoang:2017suc} 
  A.~H.~Hoang, A.~Jain, C.~Lepenik, V.~Mateu, M.~Preisser, I.~Scimemi and I.~W.~Stewart,
  arXiv:1704.01580 [hep-ph].

\bibitem{Pineda:2001zq} 
  A.~Pineda,
  JHEP {\bf 0106}, 022 (2001)
  [hep-ph/0105008].

\bibitem{Bigi:1994em}
  I.~I.~Y.~Bigi, M.~A.~Shifman, N.~G.~Uraltsev and A.~I.~Vainshtein,
  Phys.\ Rev.\ D \textbf{50}, 2234 (1994)
  [arXiv:hep-ph/9402360].

\bibitem{Beneke:1994sw}
  M.~Beneke and V.~M.~Braun,
  Nucl.\ Phys.\ \textbf{B426}, 301 (1994)
  [arXiv:hep-ph/9402364].

\bibitem{Beneke:1994rs} 
  M.~Beneke,
  Phys.\ Lett.\ B {\bf 344}, 341 (1995)
  [hep-ph/9408380].

\bibitem{Beneke:1998ui} 
  M.~Beneke,
  Phys.\ Rept.\  {\bf 317}, 1 (1999)
  [hep-ph/9807443].

\bibitem{Lee:1996yk} 
  T.~Lee,
  Phys.\ Rev.\ D {\bf 56}, 1091 (1997)
  [hep-th/9611010].

\bibitem{Bauer:2011ws} 
  C.~Bauer, G.~S.~Bali and A.~Pineda,
  Phys.\ Rev.\ Lett.\  {\bf 108}, 242002 (2012)
  [arXiv:1111.3946 [hep-ph]].
  
\bibitem{Bali:2013pla} 
  G.~S.~Bali, C.~Bauer, A.~Pineda and C.~Torrero,
  Phys.\ Rev.\ D {\bf 87}, 094517 (2013)
  [arXiv:1303.3279 [hep-lat]].
    
\bibitem{Ayala:2014yxa} 
  C.~Ayala, G.~Cvetic and A.~Pineda,
  JHEP {\bf 1409}, 045 (2014)
  [arXiv:1407.2128 [hep-ph]].
 
\bibitem{Bali:2003jq} 
  G.~S.~Bali and A.~Pineda,
  Phys.\ Rev.\ D {\bf 69}, 094001 (2004)
  [hep-ph/0310130].

\bibitem{Campanario:2005np} 
  F.~Campanario and A.~Pineda,
  Phys.\ Rev.\ D {\bf 72}, 056008 (2005)
  [hep-ph/0508217].

\bibitem{Bali:2013qla} 
  G.~S.~Bali, C.~Bauer and A.~Pineda,
  PoS LATTICE {\bf 2013}, 371 (2014)
  [arXiv:1311.0114 [hep-lat]].

\end{thebibliography}
\end{document}